\documentclass[12pt]{article}
\pdfoutput=1 
\usepackage{indentfirst}
\usepackage{natbib}
\usepackage{amsmath}
\usepackage{mathrsfs}
\usepackage{amsfonts}
\usepackage{graphicx}
\usepackage{rotating}
\usepackage{enumerate}
\usepackage{multirow}
\def\bbR{\mathbb{R}}

\makeatletter
\@addtoreset{equation}{section}\makeatother

\makeatletter
\@addtoreset{equation}{section}\makeatother

\date{}
\title{\bf An Application of Bayesian Variable Selection to Spatial Concurrent Linear Models}
\author{{\small Zuofeng Shang\thanks{Corresponding author: zshang@nd.edu}}\\
\small \textit{Environmental Change Initiative, and}\\
\small \textit{Department of Applied and Computational Mathematics and Statistics}\\
\small \textit{University of Notre Dame, Notre Dame, USA}\\
\small and \\
\small Murray K. Clayton\\
\small \textit{Department of Statistics}\\
\small \textit{University of Wisconsin-Madison, Madison, USA}\\ 
}

\begin{document}
\maketitle

\begin{abstract}
  \footnotesize{ Spatial concurrent linear models, in which the model
    coefficients are spatial processes varying at a local level, are
    flexible and useful tools for analyzing spatial data.  One
    approach places stationary Gaussian process priors on the spatial
    processes, but in applications the data may display strong
    nonstationary patterns.  In this article, we propose a Bayesian
    variable selection approach based on wavelet tools to address this
    problem.  The proposed approach does not involve any stationarity
    assumptions on the priors, and instead we impose a mixture prior
    directly on each wavelet coefficient.  We introduce an option to
    control the priors such that high resolution coefficients are more
    likely to be zero.  Computationally efficient MCMC procedures are
    provided to address posterior sampling, and uncertainty in the
    estimation is assessed through posterior means and standard
    deviations.  Examples based on simulated data demonstrate the
    estimation accuracy and advantages of the proposed method. We also
    illustrate the performance of the proposed method for real data
    obtained through remote sensing.

    \vspace{5mm} {\bf Keywords and phrases:}\,\, Bayesian estimation;
    Satellite images; Haar wavelet; Mixture prior; spike and slab prior; Blockwise Gibbs
    sampler; Inference.}
\end{abstract}

\newpage%%%%%%%%%%%%%%%%%%%%%%%%%%%%%%%%%%%%%%%%%%%%%%%%%%%%%%%%%%%%%%%%%%%%%%%%%%%%%%%%%%%%%%%%%%%%%%%%%

\section{Introduction}

\begin{figure}[htb]
\begin{center}
\includegraphics[angle=0,scale=1]{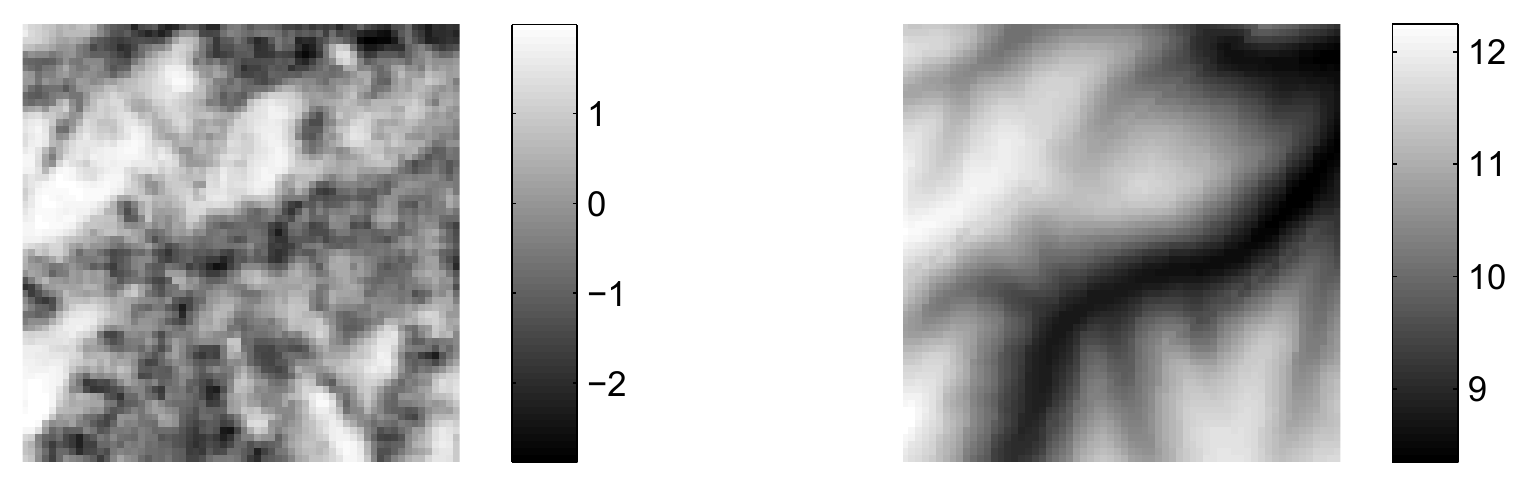}
\caption{\textit{\footnotesize Images of gypsy moth defoliation data. 
The left panel is the image of centered-and-scaled defoliation rate 
and the right panel is the image of scaled elevation. 
Low to high data values are represented by black to white tones.
The defoliation rate and elevation respectively represent the proportion
of defoliated forest and height on a per-pixel basis, and both of
the images have 30m pixel resolution. The defoliation rate data were
obtained through Landsat satellite imaging and the elevation data
were obtained by the National Elevation Data set of the US
Geological Survey.} }
\label{fig:1}
\end{center}
\end{figure}

One objective in spatial data analysis is to study the relationship
between explanatory (input) and response (output) variables through an
appropriate model.  Our interest arises when the input and output are
represented by images consisting of large numbers of pixels, as might
be obtained in remote sensing (satellite) imagery.  A particular
example consists of gypsy moth defoliation data which were obtained by
satellite from a region in the Appalachian Mountains in June-July
2006. (See Townsend \textit{et al.} (2004) for more details.)  For
these data, the response is an image representing gypsy moth
defoliation rates of oak trees.  It is of interest to relate these
rates to elevation, which can also be represented as an image (Figure
\ref{fig:1}). Several authors have observed that
defoliation rate generally increases with elevation (see, e.g.,
Kleiner and Montgomery, 1994). 

Zhang {\it et al.} (2011) assessed that relationship by using
a concurrent linear model with general form
\begin{equation}\label{svc_model}
y(\textbf{s})=A(\textbf{s})+x_1(\textbf{s})B_1(\textbf{s})+\ldots+x_K(\textbf{s})B_K(\textbf{s})+\epsilon(\textbf{s}),
\end{equation} 
where $\textbf{s}$ indicates a spatial location,
$A$ is the intercept surface, $B_1,\ldots, B_K$ are the slope surfaces,
and $\epsilon(\textbf{s})$ indicates the error term.
In the defoliation rate data (Figure \ref{fig:1}), $K=1$. 

One challenge with these data is the very large number of
observations, and the potentially large number of parameters to estimate.
Zhang {\it et al.} (2011) applied a wavelet transformation to both the  intercept and
slope surfaces 
and proposed using LASSO to estimate the model parameters.
Besides computational facility, Zhang's approach does not require the 
coefficient surfaces $A$, $B_1,\ldots,B_K$ to be stationary, and hence,
can be applied to a broad range of situations 
such as the defoliation rate data displayed 
in Figure \ref{fig:1} which appears to involve complex nonstationary patterns. 
However, it is hard to use Zhang's approach to conduct inference, 
which is the motivation of the present work. In this paper, we consider 
two major generalizations.  First, we
expand on the work of Zhang \textit{et al.} (2011) by using a Bayesian framework
based on Bayesian variable selection (BVS)
that allows for more direct inferences on the estimates.  Second,
this naturally results in a generalization of previous work on
BVS in the wavelet-based one-dimensional
time setting to a two-dimensional spatial setting.  The result is an
approach that is flexible and efficient for modeling the
relationships between image data involving complex patterns.
Furthermore, to address the large sample size and complex dependence
structure of these spatial data, we implement an efficient Gibbs sampler.
Because we reply on Zhang's modeling strategy in this paper, 
we briefly outline some notions of wavelets. We also briefly review
some previous work on BVS.

Wavelets are sets of functions whose shifts and scales form a set of basis functions. 
In particular, a bivariate wavelet consists of three functions denoted by $\varphi^r$ 
for $r=1,2,3$. When the $\varphi^r$s are chosen correctly, 
any two-dimensional square integrable function $f$ can be represented
by the following approximation, 
\begin{equation}\label{wavelet_expansion}
f(\textbf{s})\approx f_0+\sum\limits_{r=1}^3\sum\limits_{j=0}^J
\sum\limits_{k\in \Lambda_j} f_{jk}^r \varphi_{jk}^r(\textbf{s}),\,\,\textbf{s}\in [0,1)\times[0,1),
\end{equation}
where $J$ is the maximal level of decomposition,
$\varphi_{jk}^r(\textbf{s})=2^{j}\varphi^r(2^j\textbf{s}-k)$ is the
scale-and-shift transform of function $\varphi^r$, and
$\Lambda_j=\{(k_1,k_2)|\,k_1, k_2=0,1,\ldots,2^j-1\}$ is the index set
for $k$ at resolution level $j$. $\{\varphi_{jk}^r\}$ is called the
wavelet basis and $\{f_0, f_{jk}^r\}$ are the wavelet
coefficients. The transform from $f$ to $\{f_0, f_{jk}^r\}$ is called
the two-dimensional discrete wavelet transform (DWT).  If we want to
include more details or information from the image $f$, a large $J$ is
preferred, and in fact, when $J$ goes to infinity, the representation
(\ref{wavelet_expansion}) will be exact (see Daubechies 1992), which
means that all of the information on $f$ is included.  When $f$ is
locally flat, a DWT can result in a very sparse coefficient set in the
sense that most of the wavelet coefficients of $f$ are zero.

A special example is the Haar wavelet, which generates orthonormal wavelet
basis functions being constant on their supports. Using Haar wavelet,
we can express $A(\textbf{s})=W(\textbf{s})\textbf{a}$ and
$B_k(\textbf{s})=W(\textbf{s})\textbf{b}_k$, where $\textbf{a}$ and
$\textbf{b}_k$ are $d$-dimensional vectors of wavelet coefficients, and
$W(\textbf{s})$ is a row vector of length $d$ corresponding to the
Haar DWT at location $\textbf{s}$.  Note that if $J$-level wavelet
expansions are used, then $d=4^{J+1}$.  Therefore, the total number of
wavelet coefficients is $m=(K+1)d=(K+1) 4^{J+1}$. If $n$ pixles of
the image are observed, then model (\ref{svc_model}) can be rewritten as
\begin{equation}\label{con_linear_model}
\textbf{y}=X\beta+\epsilon,
\end{equation}
where $\textbf{y}=(y(\textbf{s}_1),\ldots,y(\textbf{s}_n))'$,
$X=[W,\tilde{\textbf{x}}\circ W]$ is an $n\times m$ design matrix with
``$\circ$'' denoting the Schur product, $W$ is an $n\times d$ matrix
with rows $W(\textbf{s}_i)$s,
$\beta=[\textbf{a}',\textbf{b}_1',\ldots,\textbf{b}_K']'$ is an
$m$-vector, $\tilde{\textbf{x}}=[x_k(\textbf{s}_i)]_{1\le k\le K,1\le
  i\le n}$ is an $n\times K$ matrix, and $x_k$ is the $k$-th component
of $\textbf{x}$.  In order to capture fine details, $m$ might be large. 

Next, we briefly review some references on BVS. Unless otherwise stated, 
we use $\beta_j$ for $j=1,\ldots,m$ to denote the components of $\beta$.
One version of BVS was proposed by George and McCulloch (1993), 
based on the model

(a) $\textbf{y}|\beta,\sigma^2\sim N(X\beta,\sigma^2 I)$,

(b) $\beta_j|\gamma_j \stackrel{cond.\,ind.}{\sim} (1-\gamma_j) N(0,\tau_j^2)+\gamma_j N(0,c_j \tau_j^2)$,

(c) $\gamma_j\stackrel{ind.}{\sim} \textrm{Bernoulli} (p_j)$.

where $c_j>0$, $\tau_j^2>0$ and $p_j\in (0,1)$ are fixed,
``\textit{ind.}'' means \textit{independence} and
``\textit{cond.\,ind.}'' means \textit{conditional independence}. 
Each $\gamma_j$ is a 0-1 variable and $\gamma_j$ and 0 are related 
with inclusion and exclusion of $\beta_j$ respectively when $\tau_j$s and $c_j$s 
are set at a small and a large value respectively. The authors gave procedures for
selecting $c_j$ and $\tau_j^2$  and defined the best model to be
$\hat{\gamma}=\arg\max\limits_{\gamma}\,\, p(\gamma|\textrm{data})$.
A Gibbs sampler was used for computations.

Different BVS procedures have been proposed based on variations of
(a)--(c). For instance, Smith and Kohn (1996) applied BVS to spline regression models.
They assumed that a signal
vector $\textbf{f}=(f(s_1),\ldots,f(s_n))'$ was observed with noise
and considered the model $\textbf{y}=\textbf{f}+\epsilon$, where
$\textbf{y}$ is the vector of observations and $\epsilon$ is the
vector of noise. Using spline basis expansions they rewrote this model
as $\textbf{y}=X\beta+\epsilon$, where
$\beta$ is a vector of spline coefficients and $X$ is a matrix induced by the
spline basis functions.
They proposed the following variation of (b),

(b)$'$ $\beta_j|\gamma_j \stackrel{cond.\,ind.}{\sim} (1-\gamma_j) \delta_0+\gamma_j N(0,c_j \sigma^2)$,

where $\delta_0$ is the point mass measure at zero and $c_j>0$ is
fixed.  Prior (b)$'$ is known as the spike and slab prior. 
They also developed a Gibbs sampler for computation based on
the model (a), (b)$'$, (c). Subsequently, 
Clyde \textit{et al.} (1998) and Clyde and George (2000) considered similar models
in different settings such as the one-dimensional wavelet regression problem.

Another strategy for coefficient selection was implemented for Gabor regression over the
time domain by Wolfe \textit{et
  al.} (2004). The principal difference in the Gabor approach and the wavelet approach
is that the Gabor system forms an over-complete basis whereas
the wavelet basis is complete. Wavelet approach is useful since we may choose
the wavelet basis to be orthogonal which may result in computational convenience. The
authors used Ising and Markov chain priors to model the dependence
structure among the Gabor coefficients. In order to accommodate more
flexibility, they proposed the following variations of (b)$'$ and (c),

(b)$''$ $\beta_j|\gamma_j,\tau_j^2 \stackrel{cond.\,ind.}{\sim} (1-\gamma_j) \delta_0+\gamma_j N(0,\tau_j^2)$,
\,\, $\tau_j^2\stackrel{ind.}{\sim}$ Inverse Gamma

(c)$'$ $\gamma\sim p(\gamma)$,

where $\beta_j$s denote the Gabor coefficients, 
$\tau_j^2$ may vary with $\beta_j$, 
and $p(\gamma)$ varies among the Bernoulli, 
Ising and Markov chain priors. Then, based on model (a), (b)$''$, (c)$'$,
the authors applied a Gibbs sampler to approximate the $\beta_j$s and $\tau_j^2$s. 

Other relevant references include Brown \textit{et
  al.} (2001) who used BVS based on a one-dimensional wavelet approach to
analyze curve data over time, and proposed a Metropolis-Hasting type
sampler for posterior computation.  Brown \textit{et al.} (2002)
generalized the model proposed by George and McCulloch (1993) to a
multi-dimensional situation, and proposed an estimation procedure
based on prediction.  Nott and Green (2004) discussed several
computational issues related to BVS.  Yuan and Lin (2005) explored the
relationship between LASSO and Bayesian approaches through a variable
selection view. Smith and Fahrmeir (2007) proposed a piecewise local
linear model to analyze fMRI data, and performed BVS by using Ising
priors on each local linear model. Wheeler (2009) proposed geographically
weighted LASSO to analyze spatial data. Wheeler and Waller (2009) proposed
a Bayesian framework (built upon a parametric model) analogous to ridge regression to analyze spatial 
concurrent linear model, while the proposed approach here relies on a nonparametric
wavelet approach which can capture the local behaviors of the estimates.
There are also several theoretical results on BVS including 
asymptotics of the posterior density: Jiang (2007); Jiang and Tanner
(2008), in which the authors proved density consistency under some
functional metric; and posterior model consistency: Fern\'{a}ndez
\textit{et al.} (2001); Casella \textit{et al.} (2009); Liang
\textit{et al.} (2008); Moreno \textit{et al.} (2010); and Shang and
Clayton (2011), in which the authors proved that, under suitable
conditions, the posterior probability of the true model converges to
one as the sample size grows to infinity. 

The remainder of this paper is structured as follows.  In Section 2,
two different Bayesian models will be established and the
corresponding MCMC algorithms for posterior sampling will be
described. In Section 3, simulation and real data examples
demonstrating the applications of our models and algorithms will be
provided. In particular, we discuss the matter of making inferences
for the slope and intercept surfaces.  Section 4 contains discussion,
and the supplement material contains technical details.
 
\section{Models and Algorithms}

In this section, we develop our specific modeling approach. 
To simplify the details, we only consider $K=1$ in model 
(\ref{svc_model}), i.e., only one slope surface is involved, although
generalization to multiple slope surfaces is not difficult. 
Thus, model (\ref{svc_model}) becomes the following 
model with a single covariate surface $x$
\begin{equation}\label{simple_svc_model}
y(\textbf{s}_i)=A(\textbf{s}_i)+x(\textbf{s}_i)B(\textbf{s}_i)+\epsilon(\textbf{s}_i),\,\,
i=1,\ldots,n,
\end{equation}
where $n=4^{J+2}$, $\{\textbf{s}_i\}_{i=1}^n=\{(2^{-J-2}k_1,2^{-J-2}k_2)| k_1, k_2=0,1,\ldots,2^{J+2}-1\}$
is the set of locations evenly spaced over $[0,1)\times [0,1)$,
and the $\epsilon(\textbf{s}_i)$s $\stackrel{iid.}{\sim} N(0,\sigma^2)$.
By performing a two-dimensional Haar DWT with maximal level of decomposition $J$ 
on $A$ and $B$, model (\ref{simple_svc_model})
can be written as a linear model $\textbf{y}=X\beta+\epsilon$, which is a special
case of (\ref{con_linear_model}) when $K=1$.
Here, $X$ is the $n\times m$ design matrix
induced by Haar DWT with $m=2(4^{J+1})$,
$\epsilon\sim N(\textbf{0},\sigma^2 I_n)$ is an $n$-vector of errors,
and $\beta=[\textbf{a}',\textbf{b}']'$ with $\textbf{a}$ and $\textbf{b}$
being the $(m/2)$-vectors of wavelet coefficients corresponding to surfaces $A$ and $B$.

Instead of imposing stationary prior distributions in the spatial domain of 
$A$ and $B$, we assign mixture priors in the wavelet domain 
$\beta$ corresponding to the resolution levels,
which may produce nonstationary priors for $A$ and $B$ and accommodate more
complex structures in spatial domain. 
Even if the components of $\beta$ are assumed to be {\it a priori} independent, 
when $\textbf{s}\neq\tilde{\textbf{s}}$, $A(\textbf{s})$ and $A(\tilde{\textbf{s}})$, 
$B(\textbf{s})$ and $B(\tilde{\textbf{s}})$
may still be spatially correlated. 
In fact, as $\textbf{s}$ and $\tilde{\textbf{s}}$ become 
closer in space, $A(\textbf{s})$ and $A(\tilde{\textbf{s}})$, 
$B(\textbf{s})$ and $B(\tilde{\textbf{s}})$
will share more common wavelet coefficients in their wavelet expansions, 
which makes their spatial correlations stronger.

We will consider two different Bayesian models and provide corresponding MCMC algorithms.
In both models, we assume 
\[
\textbf{y}|X,\beta,\sigma^2\sim N(X\beta,\sigma^2 I_n),\,\,\,\,\,\,\,\,\,\,\,1/\sigma^2\sim \chi_\nu^2,
\]
where $\nu$ is a fixed hyperparameter. Let $\gamma=(\gamma_1,\ldots,\gamma_m)$ with $\gamma_j$s being
the 0-1 Bernoulli variables 
indicating the exclusion and inclusion of $\beta_j$s. 
In both models we place Bernoulli priors on $\gamma$, i.e.,
$p(\gamma_1,\ldots,\gamma_m)=\prod\limits_{j=1}^m
\theta_j^{\gamma_j}(1-\theta_j)^{1-\gamma_j}$,
where $\theta_j=p(\gamma_j=1)$ is the inclusion probability. 
However, we consider different priors for $\beta$.

Our first Bayesian model requires all the nonzero components of
$\beta$ to possess a common prior variance $\tau^2$. Given $\gamma$
and $\tau^2$, the $\beta_j$s are independent with mixture priors.
\begin{eqnarray*}
\textbf{Model I:} && \beta_j|\gamma_j,\tau^2\sim (1-\gamma_j)\delta_0+\gamma_j N(0,\tau^2),\,\,\,\,\,\,\,\,\,\,\,1/\tau^2\sim \chi_\mu^2,
\end{eqnarray*}
where $\mu$ is fixed. Based on Model I, the posterior
distribution of $(\beta,\gamma,\sigma^2,\tau^2)$ is
\begin{eqnarray}\label{post:likelihood}
&&p(\beta,\gamma,\sigma^2,\tau^2|\textbf{y},X)\nonumber\\
&\propto& \left(\frac{1}{\sqrt{2\pi}\sigma}\right)^n \exp\left(-\|\textbf{y}-X\beta\|^2/(2\sigma^2)\right)\cdot\prod\limits_{j=1}^m
\left[\frac{1}{\tau}\phi\left(\frac{\beta_j}{\tau}\right)\right]^{\gamma_j}
\delta_0(\beta_j)^{1-\gamma_j}\nonumber\\
&&\cdot \frac{2^{-\nu/2}}{\Gamma(\nu/2)} \sigma^{-\nu-2}\exp(-1/(2\sigma^2))\cdot \frac{2^{-\mu/2}}{\Gamma(\mu/2)} \tau^{-\mu-2}\exp(-1/(2\tau^2)) p(\gamma),
\end{eqnarray}
where $\phi$ is the $N(0,1)$ probability density function.
If $\tau=\sigma$, then Model I is similar to one proposed by Clyde
\textit{et al.} (1998) and Li and Zhang (2010). Here we do not assume
that the variances of the coefficients are related to $\sigma$, which
makes our model flexible.  A blockwise Gibbs sampler introduced by
Godsill and Rayner (1998) and Wolfe \textit{et al.} (2004) will be
used to draw samples from the posterior distribution, as we now
describe.

{\bf Algorithm I.}\,\, Given a current state
$(\beta^{(t)},\gamma^{(t)},\sigma^{(t)}, \tau^{(t)})$.
\begin{enumerate}[(A)]
\item Update $(\gamma,\beta)$: 
\[
p(\gamma^{(t+1)}_j=1|\beta_{-j},\gamma_{-j},\sigma^{(t)},\tau^{(t)},\textbf{y},X)=\frac{1}{1+\rho_j},
\]
\[
p(\beta^{(t+1)}_j=0|\gamma^{(t+1)}_j=0,\beta_{-j},\gamma_{-j},\sigma^{(t)},\tau^{(t)},\textbf{y},X)=1,
\]
\[
\beta^{(t+1)}_j|\gamma^{(t+1)}_j=1,\beta_{-j},\gamma_{-j},\sigma^{(t)},\tau^{(t)},\textbf{y},X\sim N\left(\frac{u_j}{v_j^2},\frac{(\sigma^{(t)})^2}{v_j^2}\right),
\]
where $\gamma_{-j}=\left(\gamma^{(t+1)}_1,\ldots,\gamma^{(t+1)}_{j-1},\gamma^{(t)}_{j+1},\ldots,\gamma^{(t)}_m\right)'$,
$\beta_{-j}=\left(\beta^{(t+1)}_1,\ldots,\beta^{(t+1)}_{j-1},\beta^{(t)}_{j+1},\ldots,\beta^{(t)}_m\right)'$,
\[
u_j=(\textbf{y}-X_{-j}\beta_{-j})'X_j,\,\,\,\,v_j=\left(X_j'X_j+\frac{(\sigma^{(t)})^2}{(\tau^{(t)})^2}\right)^{1/2}
\]
with $X_j$ being the $j$-th column of $X$ and $X_{-j}=\left(X_1,\ldots,X_{j-1},X_{j+1},\ldots,X_m\right)$, and
\[
\rho_j=\frac{p(\gamma_j=0|\gamma_{-j})}{p(\gamma_j=1|\gamma_{-j})} \frac{\tau^{(t)} v_j}{\sigma^{(t)}} \exp\left(-\frac{u_j^2}{2(\sigma^{(t)})^2 v_j^2}\right).
\]
\item Update $(\sigma,\tau)$:
\[(\sigma^{(t+1)})^2|\gamma^{(t+1)},\beta^{(t+1)},\tau^{(t)},\textbf{y},X\sim IG\left(\frac{n+\nu}{2},\frac{1+\|\textbf{y}-X_{\gamma^{(t+1)}}\beta^{(t+1)}_{\gamma^{(t+1)}}\|_2^2}{2}\right),
\]
\[
(\tau^{(t+1)})^2|\gamma^{(t+1)},\beta^{(t+1)},\sigma^{(t+1)},\textbf{y},X\sim
IG\left(\frac{|\gamma^{(t+1)}|+\mu}{2},\frac{1+\|\beta^{(t+1)}\|_2^2}{2}\right),
\]
where $IG(a,b)$ denotes the inverse gamma distribution with density $g(x)\propto x^{-a-1} \exp\left(-b/x\right)$ for $x>0$.
\end{enumerate}

The derivation of Algorithm I can be found in the supplement material. Unlike the
usual non-blockwise Gibbs sampler, Algorithm I involves no matrix
inversion, and hence, is computationally efficient when $m$ is
moderate.  However, when $m$ is large, a direct application of
Algorithm I will still be time-consuming because evaluating the
quantity $u_j$ in step (A) involves intensive matrix
multiplication. To address this problem, we notice that
$V_j=\textbf{y}-X_{-j}\beta_{-j}$ and
$V_{j-1}=\textbf{y}-X_{-(j-1)}\beta_{-(j-1)}$ satisfy
\begin{equation}\label{imp:relation}
V_{j}=V_{j-1}+\beta^{(t)}_{j} X_{j}-\beta^{(t+1)}_{j-1} X_{j-1}.
\end{equation}
By (\ref{imp:relation}), $V_j$ can be obtained directly through
$V_{j-1}$, which is available from the last updating. This effectively
avoids unnecessary matrix multiplications in each iteration.  A
technique similar in spirit to (\ref{imp:relation}) to reduce the
computational burden was employed by Li and Zhang (2010), who proposed
a non-blockwise Gibbs sampler for high-dimensional structured models.

In Model I, the prior variances of the nonzero $\beta_j$s have been
set to be a common hyperparameter $\tau^2$, which seems
restrictive. Our second Bayesian model overcomes this restriction by
introducing different prior variances $\tau^2_j$s for $\beta_j$s.
Given $\gamma$ and $\tau^2_j$s, we assume the $\beta_j$s are
independent with mixture priors as follows:
\begin{eqnarray*}
\textbf{Model II:} && \beta_j|\gamma_j,\tau_j^2\sim (1-\gamma_j)\delta_0+
\gamma_j N(0,\tau_j^2),\,\,\,\,\,\,\,\,\,\,\,1/\tau_1^2,\ldots,1/\tau_m^2 \stackrel{iid.}{\sim} \chi_\mu^2,
\end{eqnarray*}
where $\mu$ is fixed.  Based on Model II, the posterior distribution of
$(\beta,\gamma,\sigma^2,\tau^2_1,\ldots,\tau^2_m)$ is
\begin{eqnarray}\label{general:post:likelihood}
&&p(\beta,\gamma,\sigma^2,\tau^2_1,\ldots,\tau^2_m|\textbf{y},X)\nonumber\\
&\propto& \left(\frac{1}{\sqrt{2\pi}\sigma}\right)^n \exp\left(-\|\textbf{y}-X\beta\|^2/(2\sigma^2)\right)\cdot
\prod\limits_{j=1}^m\left[\frac{1}{\tau_j}\phi\left(\frac{\beta_j}{\tau_j}\right)\right]^{\gamma_j}
\delta_0(\beta_j)^{1-\gamma_j}\nonumber\\
&&\cdot \frac{2^{-\nu/2}}{\Gamma(\nu/2)} \sigma^{-\nu-2}\exp(-1/(2\sigma^2))
\cdot\prod_{j=1}^m \frac{2^{-\mu/2}}{\Gamma(\mu/2)} \tau_j^{-\mu-2}\exp(-1/(2\tau_j^2)) p(\gamma),
\end{eqnarray}
where $\phi$ is the $N(0,1)$ probability density function.
Using the blockwise technique, one can draw posterior samples from
$p(\beta,\gamma,\sigma^2,\tau^2_1,\ldots,\tau^2_m|\textbf{y},X)$ with the following
algorithm:

{\bf Algorithm II.}

Given a current state
$(\beta^{(t)},\gamma^{(t)},\sigma^{(t)}, \tau^{(t)}_1,\cdots,\tau^{(t)}_m)$.
\begin{enumerate}[(A)]
\item Update $(\gamma,\beta)$: 
\[
p(\gamma^{(t+1)}_j=1|\beta_{-j},\gamma_{-j},\tau^{(t)}_1,\cdots,\tau^{(t)}_m,\sigma^{(t)},\textbf{y},X)=\frac{1}{1+\rho_j},
\]
\[
p(\beta^{(t+1)}_j=0|\gamma^{(t+1)}_j=0,\beta_{-j},\gamma_{-j},\tau^{(t)}_1,\cdots,\tau^{(t)}_m,\sigma^{(t)},\textbf{y},X)=1,
\]
\[
\beta^{(t+1)}_j|\gamma^{(t+1)}_j=1,\beta_{-j},\gamma_{-j},\tau^{(t)}_1,\cdots,\tau^{(t)}_m,\sigma^{(t)},\textbf{y},X\sim N\left(\frac{u_j}{v_j^2},\frac{(\sigma^{(t)})^2}{v_j^2}\right),
\]
where $\gamma_{-j}=\left(\gamma^{(t+1)}_1,\ldots,\gamma^{(t+1)}_{j-1},\gamma^{(t)}_{j+1},\ldots,\gamma^{(t)}_m\right)'$,
$\beta_{-j}=\left(\beta^{(t+1)}_1,\ldots,\beta^{(t+1)}_{j-1},\beta^{(t)}_{j+1},\ldots,\beta^{(t)}_m\right)'$,
\[
u_j=(\textbf{y}-X_{-j}\beta_{-j})'X_j,\,\,\,\,v_j=\left(X_j'X_j+\frac{(\sigma^{(t)})^2}{(\tau^{(t)}_j)^2}\right)^{1/2}
\]
with $X_j$ being the $j$-th column of $X$ and $X_{-j}=\left(X_1,\ldots,X_{j-1},X_{j+1},\ldots,X_m\right)$, and
\[
\rho_j=\frac{p(\gamma_j=0|\gamma_{-j})}{p(\gamma_j=1|\gamma_{-j})} \frac{\tau^{(t)}_j v_j}{\sigma^{(t)}} 
\exp\left(-\frac{u_j^2}{2(\sigma^{(t)})^2 v_j^2}\right).
\]

\item Update $\tau_j$: 
\[
(\tau^{(t+1)}_j)^2| \beta^{(t+1)}_j,\gamma^{(t+1)}_j=0,\textbf{y},X\sim 1/\chi_\mu^2,
\]
\[
(\tau^{(t+1)}_j)^2| \beta^{(t+1)}_j,\gamma^{(t+1)}_j=1,\textbf{y},X\sim IG\left(\frac{1+\mu}{2},\frac{1+(\beta^{(t+1)}_j)^2}{2}\right),
j=1,\ldots,m.
\]

\item Update $\sigma$:
\[
(\sigma^{(t+1)})^2|\gamma^{(t+1)},\beta^{(t+1)},\textbf{y},X\sim IG\left(\frac{n+\nu}{2},\frac{1+\|\textbf{y}-X_{\gamma^{(t+1)}}\beta^{(t+1)}_{\gamma^{(t+1)}}\|_2^2}{2}\right).
\]
\end{enumerate}
The derivation of Algorithm II is similar to that of Algorithm I.
Since $2m+2$ parameters have been involved in Model I, while 
$3m+1$ parameters have been involved in Model II, it takes more time to use 
Algorithm II than Algorithm I for MCMC sampling.
However, Bayesian estimates resulting from Model II may sometimes have better performance
than those resulting from Model I, which will be seen in next section. To reduce computational 
cost, a technique similar to (\ref{imp:relation}) will also be applied to Algorithm II.

\section{Numerical Results}

In this section, we apply the Bayesian methods developed in Section 2
to the concurrent linear model (\ref{simple_svc_model}) and illustrate
these methods with simulated and real datasets.
In Section 3.1, we
consider the problem of reconstructing both intercept and slope
surfaces, and use them to obtain the fitted response surface. We
assess the performance of Models I and II through four criteria:
squared bias, variance, mean square error for the estimate of the
coefficient surface, and mean square error for the
response. Comparison with the LASSO approach proposed by Zhang \textit{et al.} 
(2011) will also be demonstrated.  In Section 3.2, we
try to find the locations where the relationship between the response
and the covariate is strong. In Section 3.3, we apply our methods 
to gypsy moth defoliation data.

Let $\{\textbf{s}_i\}_{i=1}^n$ be the lattice set of locations
specified in Section 2. Denote
$\textbf{A}=(A(\textbf{s}_1),\ldots,A(\textbf{s}_n))'$ and
$\textbf{B}=(B(\textbf{s}_1),\ldots,B(\textbf{s}_n))'$.  After
obtaining the estimates $\hat{\textbf{a}}$ and $\hat{\textbf{b}}$ of
$\textbf{a}$ and $\textbf{b}$, we perform an inverse DWT to obtain the
estimates of $\textbf{A}$ and $\textbf{B}$ through
$\hat{\textbf{A}}=W\hat{\textbf{a}}$ and
$\hat{\textbf{B}}=W\hat{\textbf{b}}$, where $W\in\bbR^{n\times
  \frac{m}{2}}$ corresponds to the two-dimensional Haar DWT and
satisfies $W'W=I_{m/2}$.

The Markov chains simulated from posterior likelihoods
(\ref{post:likelihood}) and (\ref{general:post:likelihood}) will converge
quickly if the initial points of these chains are carefully
selected.  Here, we adopt an empirical procedure for this purpose.  We
first let $\hat{\beta}=(X'X)^{-1}X'\textbf{y}$ be the least squares
estimate, then we choose the initial point $\beta^{(0)}$ for the Markov
chains as a draw from $N(\hat{\beta},\tilde{\sigma}^2 I_m)$ with
$\tilde{\sigma}^2$ predetermined to be the variance of $\beta^{(0)}$. 

\subsection{Assessing the Performance of Models I and II}

We assessed the performance of Models I and II
through the numerical results by Algorithms I and II.  We chose the
true intercept surface to be
\[
A(s_1,s_2)=\left\{\begin{array}{lcr}
                              1,  & 0   \le s_1< 0.5,   0  \le s_2 < 0.5\\
                              4,  & 0.5 \le s_1< 1  ,   0  \le s_2 < 0.5\\
                              7,  & 0   \le s_1< 0.5,   0.5\le s_2 < 1  \\
                              10, & 0.5 \le s_1< 1  ,   0.5\le s_2 < 1,
            \end{array}\right.
\]
and considered two different slope surfaces: (Case I) 
\[
B(s_1,s_2)=\left\{\begin{array}{lcr}
                              1, & 0    \le s_1< 0.47, 0  \le s_2 < 0.5\\
                              3, & 0.47 \le s_1< 1,    0  \le s_2 < 0.5\\
                              5, & 0    \le s_1< 0.5,  0.5\le s_2 < 1\\
                              7, & 0.5  \le s_1< 1,    0.5\le s_2 < 1,
            \end{array}\right.
\]
and (Case II) $B(s_1, s_2)=4\sin(2\pi s_1)\cos(2\pi s_2)$, 
for $0\le s_1,s_2< 1$. 

To further explore the role played by the covariate surface, 
three covariate surfaces with different types of oscillation were
considered: 
\begin{equation}\label{X:surf1}
x_a(s_1, s_2)=4 \sin\left(4\pi (s_1+s_2)\right),
\end{equation}
\begin{equation}\label{X:surf2}
x_b(s_1, s_2)=4 \sin\left(10\pi (s_1+s_2)\right),
\end{equation}
\begin{equation}\label{X:surf3}
x_c(s_1, s_2)=4 \sin\left(15\pi (s_1+s_2)\right),\,\, 0\le s_1, s_2\le 1.
\end{equation}

We chose $J=3$ and generated data from model (\ref{simple_svc_model})
with $\sigma=1$.  Therefore, $n=1024$ and $m=512$. There are 3 nonzero
wavelet coefficients for $A$.  In Case I, $B$ is locally flat
corresponding to 3 nonzero wavelet coefficients.  (Recall that we are
using Haar wavelets.)  However, in Case II, $B$ has little local
flatness and all 256 wavelet coefficients of $B$ are nonzero.  We
fixed $\mu=\nu=6$. Let $\{a_0,a^r_{jk}| r=1,2,3, j=0,1,\ldots,J,
k\in\Lambda_j\}$ and $\{b_0,b^r_{jk}| r=1,2,3, j=0,1,\ldots,J,
k\in\Lambda_j\}$ be the components of $\textbf{a}$ and $\textbf{b}$,
and $\gamma^a_0=I(a_0\neq0)$, $\gamma^b_0=I(b_0\neq0)$,
$\gamma^a_{jkr}=I(a^r_{jk}\neq0)$, $\gamma^b_{jkr}=I(b^r_{jk}\neq0)$,
where $j$ denotes the resolution level of the wavelet coefficients and
$\Lambda_j$ denotes the collection of the indexes of the wavelet
coefficients at the $j$-th resolution level. We considered the
following three different Bernoulli priors for $\gamma$.

Prior (1):
\[
p(\gamma^a_0=1)=p(\gamma^b_0=1)=0.5,\,\, p(\gamma^a_{jkr}=1)=p(\gamma^b_{jkr}=1)=0.5\phi^j,\, r=1,2,3, j=0,\ldots,J,\,\, k\in \Lambda_j. 
\]
Prior (2):
\[
p(\gamma^a_0=1)=p(\gamma^b_0=1)=0.5,\,\, p(\gamma^a_{jkr}=1)=0.5\phi^{j}, p(\gamma^b_{jkr}=1)=0.5,\,r=1,2,3, j=0,\ldots,J,\,\, k\in \Lambda_j.
\]
Prior (3):
\[
p(\gamma^a_0=1)=p(\gamma^b_0=1)=0.5,\,\, p(\gamma^a_{jkr}=1)=0.5\phi^{8j}, p(\gamma^b_{jkr}=1)=0.5,\,r=1,2,3, j=0,\ldots,J,\,\, k\in \Lambda_j.
\]

Different $\phi$ values and the resultant Bernoulli priors can
produce difference levels of sparsity in the estimates. Thus, the selection of $\phi$ is
purely empirical depending on how mush sparsity is expected in the estimates. For
instance, if a practitioner expects that the estimate should be fairly sparse, then
one can choose  to be relatively smaller such as $\phi = 0.7$; otherwise, one may just
use $\phi = 0.9$ to produce certain amount of sparsity or even use $\phi = 1$ to fully let
the model drive the amount of sparsity in the estimates since $\phi = 1$ corresponds
to indifference Bernoulli prior for the coefficients

We considered $\phi=1,0.9,0.8,0.7$. Note that when $\phi=1$, Priors (1)--(3) all become indifference priors. 
We applied Prior (1) to Case I, and 
applied Priors (2) and (3) to Case II. Prior (1) puts 
smaller weights on the higher level wavelet coefficients of both 
surfaces $A$ and $B$ so that they have larger prior probability 
to be zero, while Priors (2) and (3) only do this for surface $A$ 
but assign neutral probabilities to the wavelet coefficients of surface $B$.
 
For each of the covariate surfaces (\ref{X:surf1})--(\ref{X:surf3}) 
and for both Cases I and II, we repeated the simulations $L=50$ times. 
For the $l$-th replication with $l=1,\ldots,L$, Markov chains 
with length 5000 were generated from the posterior distribution (\ref{post:likelihood}), 
and the first 2500 served as 
burn-ins. Gelman-Rubin's factors (see Gelman \textit{et al.}, 2003) for all chains were below 1.1, 
suggesting that all chains converged well. The estimates 
$\hat{A}^l$ and $\hat{B}^l$ of $A$ and $B$ based on the $l$-th replication
were obtained through averaging the last 2500 posterior samples.

To assess performance, we borrowed an idea from Fan \textit{et al}. (2010) to calculate 
the squared bias, variance and mean square errors of the estimates. 
To state our method, we let $\hat{A}^l_i$, $\hat{B}^l_i$, $A_i$ 
and $B_i$ be the values of $\hat{A}^l$, $\hat{B}^l$, $A$ and $B$ 
at pixel $\tilde{\textbf{s}}_i$ with $\{\tilde{\textbf{s}}_i\}=\{(s_1/100,s_2/100)| s_1,s_2=0,1,\ldots,99\}$ being
the $100\times 100$ uniform grid of pixels over $[0,1)\times [0,1)$. 
Thus, there are $N=10^4$ pixels being evaluated. Note that 
$\{\tilde{\textbf{s}}_i\}$ have been chosen
to be different from the locations where data were drawn for the purposes
of assessing the performance of the estimates at new locations.
We define the average squared bias to be
\[
Bias_A^2=\frac{1}{N}\sum\limits_{i=1}^{N}\left(\sum\limits_{l=1}^{L}\frac{\hat{A}_i^l-A_i}{L}\right)^2,
\]
\[
Bias_B^2=\frac{1}{N}\sum\limits_{i=1}^{N}\left(\sum\limits_{l=1}^{L}\frac{\hat{B}_i^l-B_i}{L}\right)^2,
\]
and define the average variance to be
\[
Var_A=\frac{1}{N}\sum\limits_{i=1}^N \sum\limits_{l=1}^{L}\left(\hat{A}_i^l-\frac{1}{L}\sum\limits_{l=1}^L \hat{A}_i^l\right)^2/L,
\]
\[
Var_B=\frac{1}{N}\sum\limits_{i=1}^N \sum\limits_{l=1}^{L}\left(\hat{B}_i^l-\frac{1}{L}\sum\limits_{l=1}^L \hat{B}_i^l\right)^2/L.
\]
The average mean square errors for $A$, $B$ are then defined to be
$MSE_A=Bias_A^2+Var_A$ and $MSE_B=Bias_B^2+Var_B$. The average mean
square error for the response is defined to be
$MSE_y=\sum\limits_{i=1}^N
\sum\limits_{l=1}^L\left(\hat{A}_i^l+x_i\hat{B}_i^l-(A_i+x_i
  B_i)\right)^2/(NL)$, where $x_i=x(\tilde{\textbf{s}}_i)$.

We first assessed the performance of Model I with Algorithm I.  Tables
\ref{table:BVMI} and \ref{table:BVMII} summarize the average squared
bias, variance and mean square error of the estimates by using both
Algorithm I and LASSO.  Since Priors (2) and (3) coincide with each
other when $\phi=1$, we only recorded the results corresponding to
Prior (2) when $\phi=1$.  Several findings result from these
tables. First, for Case I where both $A$ and $B$ are piecewise
constant, the Bayesian estimates corresponding to all the covariate
surfaces $x_a$, $x_b$ and $x_c$ have similar performance in terms of
$MSE_A$, $MSE_B$ and $MSE_y$.  For estimating $A$, the Bayesian method
results in smaller mean square errors than LASSO, while for estimating
$B$, the Bayesian and LASSO methods result in comparable mean square
errors. Second, for Case II where $A$ is piecewise constant but $B$ is
smooth, the Bayesian estimates corresponding to $x_c$ are slightly
better than those corresponding to $x_a$ and $x_b$ in terms of $MSE_A$
and $MSE_B$.  Zhang \textit{et al.} (2011) observed similar effects of
the covariate surfaces on the LASSO estimates.  We can also see that,
for $\phi=0.9, 0.8, 0.7$, Prior (3) results in smaller $MSE_A$ than
Prior (2). Compared with LASSO, the Bayesian approach corresponding to
Prior (3) produces smaller $MSE_A$, but produces slightly larger
$MSE_B$. Third, for both Priors (2) and (3), when $\phi$ decreases,
the average variances of the posterior estimates of both $A$ and $B$
decrease.

\begin{table}[ptbh]
\begin{center}
\scalebox{0.7}{%
\begin{tabular}{c|cc|ccccccc} \hline
Surface                &\multicolumn{2}{c|}{Method}   & $Bias_A^2$   & $Bias_B^2$   & $Var_A$  & $Var_B$   & $MSE_A$  & $MSE_B$ & $MSE_y$\\ \hline
\multirow{5}{*}{$x_a$} &$\phi$           &$=1$        & 0.0004       & 0.0600       & 0.0146   & 0.0010    & 0.0149   & 0.0610  & 0.0223 \\
                       &                 &$=0.9$      & 0.0002       & 0.0600       & 0.0091   & 0.0007    & 0.0093   & 0.0607  & 0.0150 \\
                       &                 &$=0.8$      & 0.0002       & 0.0601       & 0.0072   & 0.0006    & 0.0074   & 0.0606  & 0.0115 \\ 
                       &                 &$=0.7$      & 0.0002       & 0.0601       & 0.0054   & 0.0005    & 0.0056   & 0.0605  & 0.0094 \\ 
\cline{2-10}                       
                       &\multicolumn{2}{c|}{LASSO}    & 0.0389       & 0.0599       & 0.0038   & 0.0088    & 0.0427   & 0.0687  & 0.0864\\ \hline  

\multirow{5}{*}{$x_b$} &$\phi$           &$=1$        & 0.0002       & 0.0601       & 0.0132   & 0.0008    & 0.0134   & 0.0609  & 0.0209 \\
                       &                 &$=0.9$      & 0.0001       & 0.0601       & 0.0086   & 0.0006    & 0.0087   & 0.0607  & 0.0140 \\
                       &                 &$=0.8$      & 0.0001       & 0.0601       & 0.0064   & 0.0005    & 0.0065   & 0.0606  & 0.0108 \\    
                       &                 &$=0.7$      & 0.0001       & 0.0601       & 0.0051   & 0.0004    & 0.0052   & 0.0605  & 0.0089 \\ 
\cline{2-10}                       
                       &\multicolumn{2}{c|}{LASSO}    & 0.0312       & 0.0595       & 0.0034   & 0.0061    & 0.0346   & 0.0656  & 0.0754 \\ \hline           
\multirow{5}{*}{$x_c$} &$\phi$           &$=1$        & 0.0002       & 0.0603       & 0.0131   & 0.0010    & 0.0133   & 0.0613  & 0.0216 \\
                       &                 &$=0.9$      & 0.0001       & 0.0602       & 0.0082   & 0.0007    & 0.0083   & 0.0610  & 0.0145 \\
                       &                 &$=0.8$      & 0.0001       & 0.0602       & 0.0060   & 0.0006    & 0.0061   & 0.0608  & 0.0111 \\
                       &                 &$=0.7$      & 0.0001       & 0.0602       & 0.0046   & 0.0005    & 0.0047   & 0.0607  & 0.0090 \\ 
\cline{2-10}                       
                       &\multicolumn{2}{c|}{LASSO}    & 0.0341       & 0.0602       & 0.0038   & 0.0044    & 0.0379   & 0.0646  & 0.0731 \\ \hline     
\end{tabular}}
\end{center}
\caption{\textit{\footnotesize Average squared bias, variance and mean 
square error related to Case I when Bayesian and LASSO approaches 
have been applied. For the Bayesian approach, Model I with Algorithm I
has been implemented and Prior (1) has been imposed on the vector of Bernoulli variables $\gamma$.}}
\label{table:BVMI}
\end{table}

\begin{table}[ptbh]
\begin{center}
\scalebox{0.7}{%
\begin{tabular}{c|c|cc|ccccccc} \hline
Surface                &\multicolumn{2}{c}{Method}       &       &$Bias_A^2$ & $Bias_B^2$ & $Var_A$ & $Var_B$ & $MSE_A$ & $MSE_B$ & $MSE_y$  \\ \hline
\multirow{8}{*}{$x_a$} &\multirow{4}{*}{Prior (2)}&$\phi$& $=1$  & 0.8216    & 0.3064     & 0.6070  & 0.0768  & 1.4286  & 0.3832  & 0.9386   \\
                       &                          &      & $=0.9$& 0.3351    & 0.2641     & 0.3963  & 0.0591  & 0.7314  & 0.3232  & 0.9582   \\
                       &                          &      & $=0.8$& 0.1473    & 0.2479     & 0.1629  & 0.0400  & 0.3103  & 0.2879  & 0.9732   \\
                       &                          &      & $=0.7$& 0.0828    & 0.2429     & 0.1196  & 0.0370  & 0.2023  & 0.2799  & 0.9872   \\
\cline{2-11}           
                       &\multirow{3}{*}{Prior (3)}&$\phi$& $=0.9$& 0.0237    & 0.2403     & 0.0336  & 0.0298  & 0.0573  & 0.2702  & 1.0124   \\
                       &                          &      & $=0.8$& 0.0144    & 0.2405     & 0.0139  & 0.0285  & 0.0283  & 0.2691  & 1.0281   \\
                       &                          &      & $=0.7$& 0.0137    & 0.2411     & 0.0121  & 0.0284  & 0.0259  & 0.2695  & 1.0305   \\ 
\cline{2-11}
                       &\multicolumn{2}{c}{LASSO}        &       & 0.1298    & 0.1987     & 0.0222  & 0.0355  & 0.1520  & 0.2342  & 0.8599   \\ \hline
\multirow{8}{*}{$x_b$} &\multirow{4}{*}{Prior (2)}&$\phi$& $=1$  & 0.0952    & 0.2069     & 0.1149  & 0.0318  & 0.2101  & 0.2387  & 0.9641   \\
                       &                          &      & $=0.9$& 0.0691    & 0.2037     & 0.0817  & 0.0304  & 0.1507  & 0.2341  & 0.9696   \\
                       &                          &      & $=0.8$& 0.0535    & 0.2034     & 0.0596  & 0.0294  & 0.1131  & 0.2327  & 0.9789   \\
                       &                          &      & $=0.7$& 0.0440    & 0.2033     & 0.0440  & 0.0286  & 0.0880  & 0.2319  & 0.9879   \\ 
\cline{2-11}
                       &\multirow{3}{*}{Prior (3)}&$\phi$& $=0.9$& 0.0313    & 0.2060     & 0.0166  & 0.0269  & 0.0479  & 0.2329  & 1.0134   \\                             &                          &      & $=0.8$& 0.0288    & 0.2065     & 0.0076  & 0.0267  & 0.0364  & 0.2332  & 1.0270   \\    
                       &                          &      & $=0.7$& 0.0285    & 0.2068     & 0.0061  & 0.0268  & 0.0346  & 0.2336  & 1.0302   \\ 
\cline{2-11}                        
                       &\multicolumn{2}{c}{LASSO}        &       & 0.1212    & 0.1885     & 0.0067  & 0.0237  & 0.1279  & 0.2122  & 0.9374   \\ \hline
\multirow{8}{*}{$x_c$} &\multirow{4}{*}{Prior (2)}&$\phi$& $=1$  & 0.0427    & 0.1994     & 0.0707  & 0.0286  & 0.1134  & 0.2280  & 1.0131   \\
                       &                          &      & $=0.9$& 0.0248    & 0.1986     & 0.0486  & 0.0271  & 0.0734  & 0.2257  & 1.0271   \\
                       &                          &      & $=0.8$& 0.0149    & 0.1990     & 0.0345  & 0.0265  & 0.0494  & 0.2255  & 1.0428   \\
                       &                          &      & $=0.7$& 0.0090    & 0.1993     & 0.0243  & 0.0261  & 0.0333  & 0.2253  & 1.0569   \\ 
\cline{2-11}
                       &\multirow{3}{*}{Prior (3)}&$\phi$& $=0.9$& 0.0025    & 0.1997     & 0.0088  & 0.0263  & 0.0113  & 0.2260  & 1.0893   \\
                       &                          &      & $=0.8$& 0.0015    & 0.1999     & 0.0032  & 0.0260  & 0.0047  & 0.2259  & 1.1052   \\
                       &                          &      & $=0.7$& 0.0014    & 0.1999     & 0.0029  & 0.0261  & 0.0043  & 0.2259  & 1.1075   \\ 
\cline{2-11}
                       &\multicolumn{2}{c}{LASSO}        &       & 0.0549    & 0.1941     & 0.0039  & 0.0199  & 0.0588  & 0.2139  & 1.2012   \\ \hline     
\end{tabular}}
\end{center}
\caption{\textit{\footnotesize Average squared bias, variance and mean 
square error related to Case II when Bayesian and LASSO approaches 
have been applied. For the Bayesian approach, Model I with Algorithm I
has been implemented and Priors (2) and (3) have been imposed on the vector of Bernoulli variables $\gamma$.}}
\label{table:BVMII}
\end{table}

To examine Model II with Algorithm II, we repeated the
simulations 50 times and each time generated 5000 MCMC samples based
on the posterior distribution (\ref{general:post:likelihood}).  We
then treated the first half as burn-ins. Convergence was monitored
through Gelman-Rubin's factors. Tables \ref{table:BVMIII} and
\ref{table:BVMIV} summarize the results of using Algorithm II.
Comparing Tables \ref{table:BVMI} and \ref{table:BVMIII}, and Tables
\ref{table:BVMII} and \ref{table:BVMIV}, two observations can be made:
(1) for Case I in which $B$ is piecewise constant, Model I and Model
II result in comparable $MSE_A$ and $MSE_B$, while Model I corresponds
to slightly smaller $MSE_y$; (2) for Case II in which $B$ is smooth,
Model II outperforms Model I in terms of $MSE_A$, $MSE_B$ and $MSE_y$.

\begin{table}[ptbh]
\begin{center}
\scalebox{0.7}{%
\begin{tabular}{c|cc|ccccccc} \hline
Surface                &\multicolumn{2}{c|}{Method}   & $Bias_A^2$   & $Bias_B^2$   & $Var_A$  & $Var_B$   & $MSE_A$  & $MSE_B$ & $MSE_y$\\ \hline
\multirow{4}{*}{$x_a$} &$\phi$           &$=1$        & 0.0003       & 0.0604       & 0.0132   & 0.0067    & 0.0135   & 0.0670  & 0.0863 \\
                       &                 &$=0.9$      & 0.0002       & 0.0602       & 0.0103   & 0.0046    & 0.0105   & 0.0648  & 0.0639 \\
                       &                 &$=0.8$      & 0.0002       & 0.0601       & 0.0079   & 0.0033    & 0.0080   & 0.0633  & 0.0466 \\ 
                       &                 &$=0.7$      & 0.0001       & 0.0600       & 0.0061   & 0.0023    & 0.0062   & 0.0623  & 0.0333 \\ \hline
\multirow{4}{*}{$x_b$} &$\phi$           &$=1$        & 0.0006       & 0.0601       & 0.0188   & 0.0075    & 0.0194   & 0.0676  & 0.0909 \\
                       &                 &$=0.9$      & 0.0004       & 0.0601       & 0.0137   & 0.0054    & 0.0141   & 0.0655  & 0.0669 \\
                       &                 &$=0.8$      & 0.0003       & 0.0601       & 0.0100   & 0.0039    & 0.0103   & 0.0640  & 0.0486 \\    
                       &                 &$=0.7$      & 0.0002       & 0.0601       & 0.0073   & 0.0028    & 0.0075   & 0.0628  & 0.0346 \\ \hline   \multirow{4}{*}{$x_c$}&$\phi$            &$=1$        & 0.0004       & 0.0598       & 0.0217   & 0.0082    & 0.0221   & 0.0680  & 0.0933   \\
                       &                 &$=0.9$      & 0.0003       & 0.0598       & 0.0152   & 0.0059    & 0.0155   & 0.0657  & 0.0680   \\
                       &                 &$=0.8$      & 0.0002       & 0.0598       & 0.0106   & 0.0042    & 0.0108   & 0.0640  & 0.0488 \\
                       &                 &$=0.7$      & 0.0001       & 0.0598       & 0.0075   & 0.0030    & 0.0076   & 0.0628  & 0.0345 \\ \hline
\end{tabular}}
\end{center}
\caption{\textit{\footnotesize Average squared bias, variance and 
mean square error related to Case I when a Bayesian approach has 
been applied. Model II with Algorithm II
has been implemented and Prior (1) has been imposed on the vector of Bernoulli variables $\gamma$.}}
\label{table:BVMIII}
\end{table}

\begin{table}[ptbh]
\begin{center}
\scalebox{0.7}{%
\begin{tabular}{c|c|cc|ccccccc} \hline
Surface                &\multicolumn{2}{c}{Method}       &       &$Bias_A^2$ & $Bias_B^2$ & $Var_A$ & $Var_B$ & $MSE_A$ & $MSE_B$ & $MSE_y$  \\ \hline
\multirow{7}{*}{$x_a$} &\multirow{4}{*}{Prior (2)}&$\phi$& $=1$  & 0.0510    & 0.2212     & 0.0222  & 0.0232  & 0.0733  & 0.2444  & 0.9304   \\
                       &                          &      & $=0.9$& 0.0451    & 0.2187     & 0.0206  & 0.0234  & 0.0657  & 0.2421  & 0.9356   \\
                       &                          &      & $=0.8$& 0.0382    & 0.2168     & 0.0187  & 0.0236  & 0.0569  & 0.2404  & 0.9398   \\
                       &                          &      & $=0.7$& 0.0342    & 0.2151     & 0.0181  & 0.0238  & 0.0523  & 0.2389  & 0.9436   \\
\cline{2-11}           
                       &\multirow{3}{*}{Prior (3)}&$\phi$& $=0.9$& 0.0221    & 0.2109     & 0.0168  & 0.0251  & 0.0389  & 0.2360  & 0.9500   \\
                       &                          &      & $=0.8$& 0.0199    & 0.2102     & 0.0161  & 0.0253  & 0.0360  & 0.2355  & 0.9526   \\
                       &                          &      & $=0.7$& 0.0203    & 0.2103     & 0.0163  & 0.0253  & 0.0366  & 0.2356  & 0.9531   \\ \hline
\multirow{7}{*}{$x_b$} &\multirow{4}{*}{Prior (2)}&$\phi$& $=1$  & 0.0200    & 0.1846     & 0.0112  & 0.0218  & 0.0312  & 0.2064  & 0.9020   \\
                       &                          &      & $=0.9$& 0.0162    & 0.1852     & 0.0088  & 0.0213  & 0.0250  & 0.2065  & 0.9035   \\
                       &                          &      & $=0.8$& 0.0143    & 0.1848     & 0.0072  & 0.0214  & 0.0215  & 0.2062  & 0.9080   \\
                       &                          &      & $=0.7$& 0.0134    & 0.1845     & 0.0060  & 0.0215  & 0.0194  & 0.2060  & 0.9115   \\ 
\cline{2-11}
                       &\multirow{3}{*}{Prior (3)}&$\phi$& $=0.9$& 0.0133    & 0.1828     & 0.0048  & 0.0222  & 0.0181  & 0.2050  & 0.9232   \\                             &                          &      & $=0.8$& 0.0134    & 0.1828     & 0.0046  & 0.0223  & 0.0180  & 0.2051  & 0.9270   \\    
                       &                          &      & $=0.7$& 0.0133    & 0.1829     & 0.0046  & 0.0223  & 0.0179  & 0.2052  & 0.9280   \\ \hline
\multirow{7}{*}{$x_c$} &\multirow{4}{*}{Prior (2)}&$\phi$& $=1$  & 0.0093    & 0.1811     & 0.0114  & 0.0203  & 0.0207  & 0.2014  & 0.9373   \\
                       &                          &      & $=0.9$& 0.0052    & 0.1808     & 0.0085  & 0.0204  & 0.0138  & 0.2013  & 0.9551   \\
                       &                          &      & $=0.8$& 0.0029    & 0.1808     & 0.0065  & 0.0204  & 0.0095  & 0.2012  & 0.9663   \\
                       &                          &      & $=0.7$& 0.0015    & 0.1807     & 0.0052  & 0.0203  & 0.0067  & 0.2010  & 0.9760   \\ 
\cline{2-11}
                       &\multirow{3}{*}{Prior (3)}&$\phi$& $=0.9$& 0.0002    & 0.1810     & 0.0033  & 0.0203  & 0.0035  & 0.2013  & 0.9920   \\
                       &                          &      & $=0.8$& 0.0001    & 0.1810     & 0.0031  & 0.0203  & 0.0032  & 0.2013  & 1.0004   \\
                       &                          &      & $=0.7$& 0.0001    & 0.1810     & 0.0031  & 0.0203  & 0.0032  & 0.2013  & 1.0010   \\ \hline     
\end{tabular}}
\end{center}
\caption{\textit{\footnotesize Average squared bias, variance and 
mean square error related to Case II when a Bayesian approach
has been applied. Model II with Algorithm II has been implemented and 
Priors (2) and (3) have been imposed on the vector of Bernoulli variables $\gamma$.}}
\label{table:BVMIV}
\end{table}

\subsection{Detecting Where the Slopes Are Nonzero}

Our modeling approach allows for nonstationarity in the $B$ surface, 
and in particular, it is possible that the
relationship between the $y$ and $x$ surfaces vary over
space.  Therefore, it is of interest to detect the regions where the
response has a strong relationship with the covariate. This is
equivalent to detecting the locations or pixels on the image where the
slopes deviate from zero.  To accomplish this, we construct a
$100(1-\alpha)\%$ credible interval for $B(\textbf{s})$ at each pixel
$\textbf{s}$. If the credible interval at $\textbf{s}$ excludes zero,
then that gives evidence that $B(\textbf{s})$ deviates from zero. Note
that the upper and lower bounds of all the credible intervals form
two-dimensional surfaces which together we call an uncertainty
band. 
Unlike one-dimensional wavelet regression problem
where the graphical demonstration of uncertainty bands is feasible
(see, e.g., Chipman \textit{et al.} 1997), it is difficult to
effectively plot the two-dimensional uncertainty bands.  In this
section, we use an alternative method to address this difficulty.
Before proceeding further, we perform some useful calculations.

We denote $B_i=B(\textbf{s}_i)$, and let
$\textbf{b}^{(1)},\ldots,\textbf{b}^{(T)}$ be $T$ posterior samples of
$\textbf{b}$, where $\textbf{b}$ denotes the vector of wavelet
coefficients of the surface $B$.  Let
$B_i^{(t)}=W(\textbf{s}_i)\textbf{b}^{(t)}$ for $t=1,\ldots,T$.  The
Bayesian estimate of $B_i$ is
\[
\hat{B}_i=\sum\limits_{t=1}^{T}B_i^{(t)}/T=W(\textbf{s}_i)\hat{\textbf{b}},
\] 
where $\hat{\textbf{b}}=\sum\limits_{t=1}^{T}\textbf{b}^{(t)}/T$.  The
posterior variance of $B_i^{(t)}$, $t=1,\ldots,T$, is
\[
\hat{\sigma}_i^2=W(\textbf{s}_i) \hat{\Sigma} W(\textbf{s}_i)',
\]
where $\hat{\Sigma}=\sum\limits_{t=1}^{T}
\left(\textbf{b}^{(t)}-\hat{\textbf{b}}\right)\left(\textbf{b}^{(t)}-\hat{\textbf{b}}\right)'/(T-1)$
is an $m\times m$ matrix. We call $\hat{\sigma}_i$ the posterior
standard deviation (PSD) of $B$ at pixel $\textbf{s}_i$.

We find the pixels at which the slopes deviate from zero, and also
classify the pixels according to the magnitudes and signs of the
slopes.  For this purpose, we construct a choropleth map to indicate
$\hat{B}_i\ge \Delta$, $0\le \hat{B}_i<\Delta$, $-\Delta<\hat{B}_i<0$
and $\hat{B}_i\le -\Delta$, with $\Delta>0$ a suitably selected
threshold.

In the simulated and real data examples discussed later, a majority of
the posterior distributions $p(B_i|\textbf{y},X)$ of $B_i$ are
unimodal and roughly symmetric. Therefore, it is convenient to
approximate $p(B_i|\textbf{y},X)$ by a normal distribution with center
and scale being $\hat{B}_i$ and $\hat{\sigma}_i$.  Using an analogy to
the concept of frequentist $p$-value, if
$|\hat{B}_i/\hat{\sigma}_i|>1.96$, then we believe with strong
evidence that $B_i\neq 0$ and represent this situation by $p<0.05$; if
$1.64\le|\hat{B}_i/\hat{\sigma}_i|\le 1.96$, then we believe with
moderate evidence that $B_i\neq0$ and represent this situation by
$0.05\le p \le 0.1$; otherwise, we believe that $B_i$ might be close
to zero and represent this situation by $p>0.1$. Note that this is
analogous to the interpretation of a frequentist $p$-value.  In a
choropleth map, we designate the various possibilities for $p$ by
different using different line-patterns.

In a simulation study, the $A$ surface was defined as in Section 3.1
and the $B$ surface was defined by Case II in Section 3.1, i.e.,
$B(s_1,s_2)=4 \sin(2\pi s_1)\cos(2\pi s_2),\,\, 0\le s_1,s_2\le 1$.
Note that $B$ is smooth with zero values at some pixels.  Algorithm I
under Model I was implemented, and we set $\Delta=2$ which is half of
the maximum value of $|B|$.

In addition, we chose $J=4$ and generated data from model
(\ref{simple_svc_model}) with $\sigma=1$.  Thus, $n=4096$ and
$m=2048$.  We chose the hyperparameters $\mu=\nu=6$ and prior (3)
defined in Section 3.1 was used for the Bernoulli variable $\gamma$
for each of the cases $\phi=1, 0.9, 0.8, 0.7$.  Markov chains of
length 5000 were simulated with the first half burn-ins, and we used
the second half for calculations.  Convergence was assessed through
Gelman-Rubin's factors.  

\begin{figure}[htb]
\begin{center}
\includegraphics[height=40cm, width=32cm, angle=0,scale=0.5]{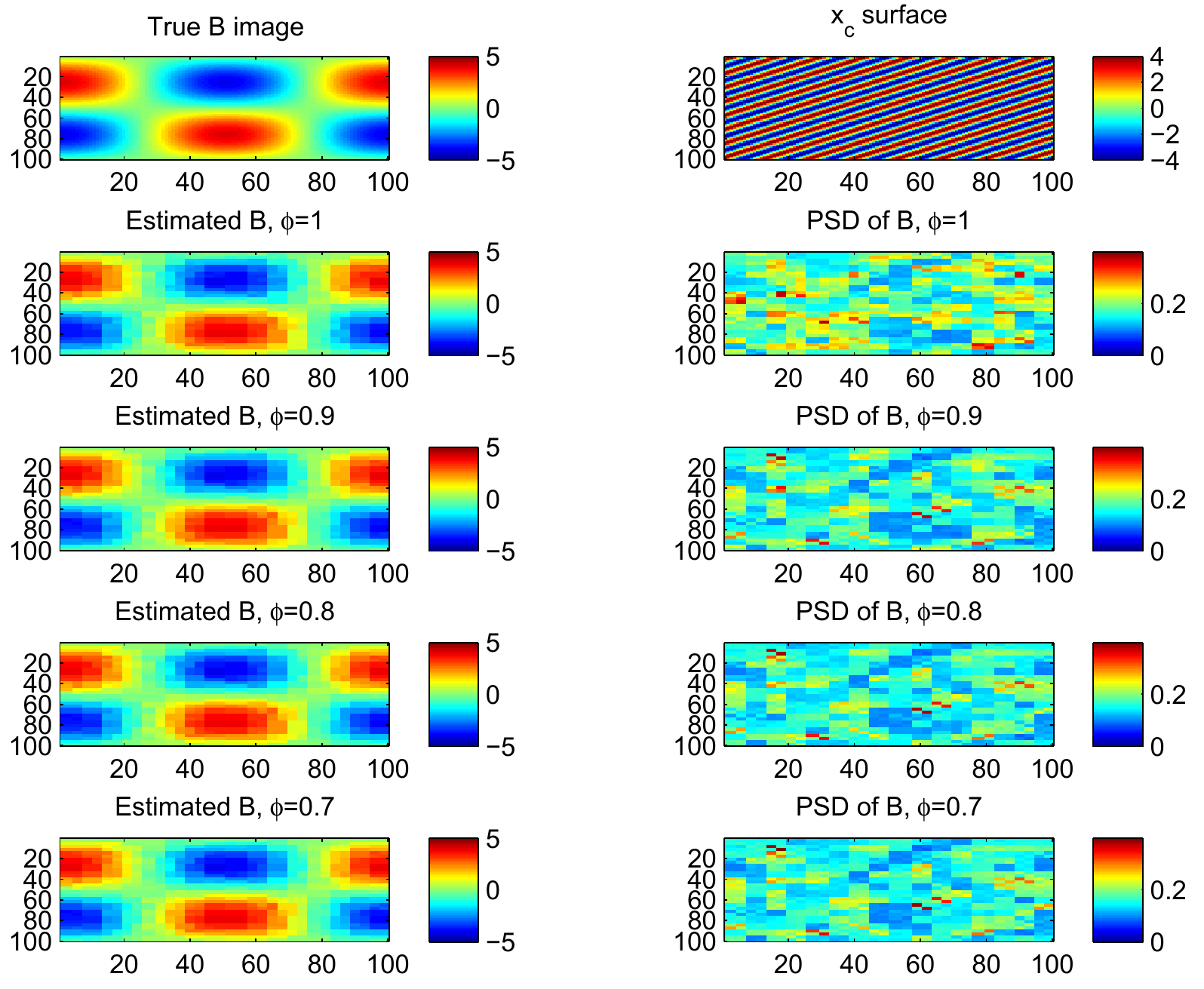}
\caption{\textit{\footnotesize Section 3.2. Images of estimated $B$ and the PSD of $B$ for 
$\phi=1,0.9,0.8,0.7$. Covariate surface $x_c$ and Prior (3) were used. Images of the true $B$ and $x_c$
are also shown.}}
\label{fig:BSigmaX3}
\end{center}
\end{figure}

Figure \ref{fig:BSigmaX3} displays the images of $\hat{B}$ and the PSD
of $B$ corresponding to $\phi=1, 0.9, 0.8, 0.7$ when using $x_c$
defined in Section 4.1 as the covariate surface.  We observe that all
the $\hat{B}$ images graphically resemble the true $B$, and the PSD of
$B$ for $\phi=1$ appear to be greater than those for
$\phi=0.9,0.8,0.7$. We also observe that when $\phi$ decreases, the
$\hat{B}$ images become slightly sparser in the sense that larger
square regions appear on the images. This is because when the
Bernoulli probabilities associated with higher level wavelet
coefficients become smaller, the finer details will be dropped and the
basis supports with smaller sizes will merge into larger square
regions.

\begin{figure}[htb]
\begin{center}
\includegraphics[height=40cm, width=32cm, angle=0,scale=0.4]{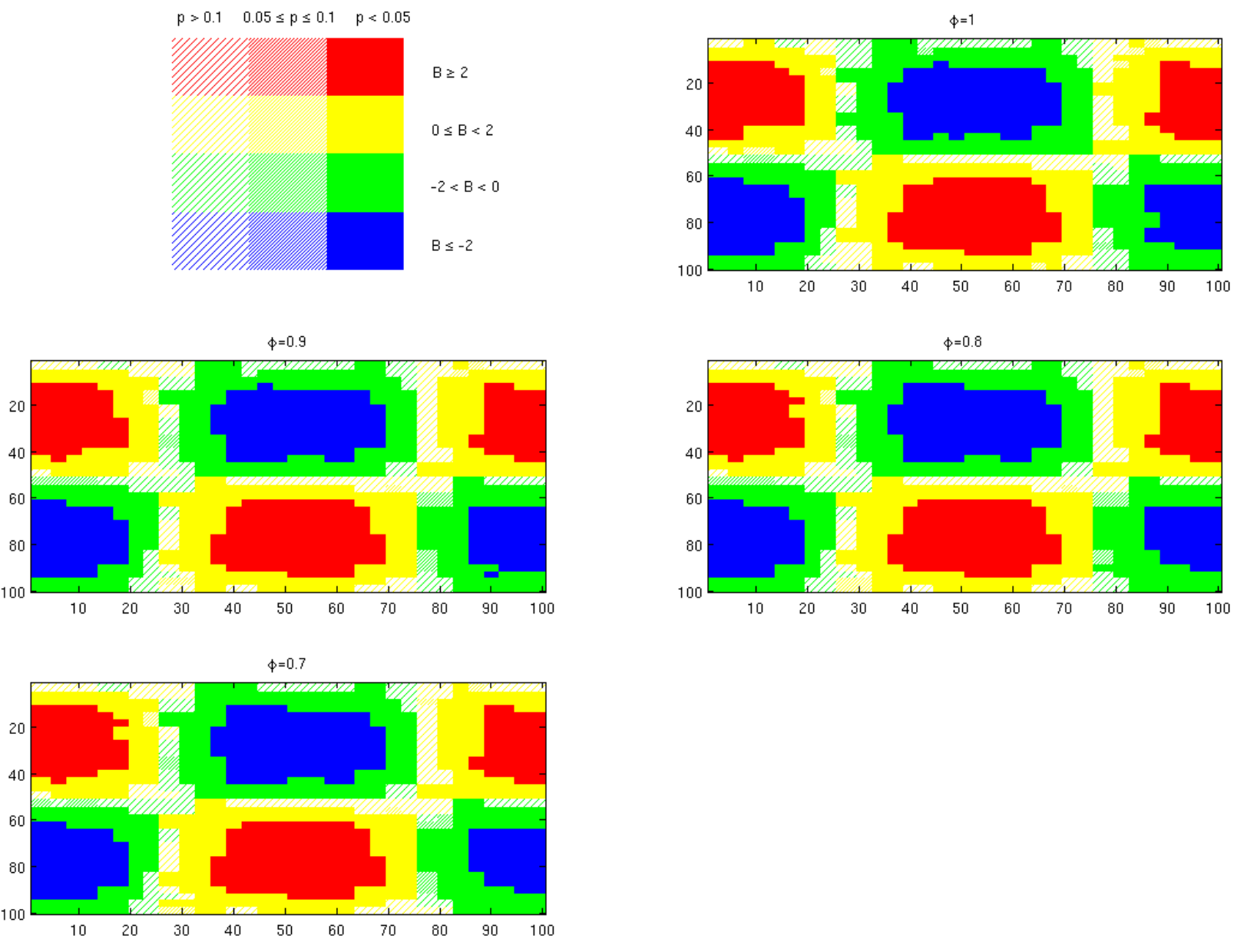}
\caption{\textit{\footnotesize Section 3.2. Choropleth map of $\hat{B}$ for $\phi=1, 0.9, 0.8, 0.7$. 
Covariate surface $x_c$ and Prior (3) were used.
Red indicates $\hat{B}_i\ge 2$; Yellow: $0\le\hat{B}_i<2$; 
Green: $-2<\hat{B}_i<0$; Blue: $\hat{B}_i\le-2$. 
Filled boxes: $p<0.05$; 
boxes with dense lines: $0.05\le p\le 0.1$; 
boxes with thin lines: $p>0.1$.}}
\label{fig:ClassX3}
\end{center}
\end{figure}

As displayed in Figure \ref{fig:BSigmaX3}, there are three peaks
(indicated by red) and three valleys (indicated by blue) regularly
arranged on the true $B$ image, and the values of the true $B$ at the
pixels around the peaks and valleys deviate from zero, while they are
close to zero elsewhere. Figure \ref{fig:ClassX3} displays the
choropleth map for $\hat{B}$ corresponding to various $\phi$
values. We observe that the locations where the $B$ values deviate
from zero are correctly detected and changing $\phi$ makes little
change in the detection results.

\subsection{Applications to Gypsy Moth Defoliation Data}

We next use the proposed Bayesian approach to analyze the gypsy moth
defoliation data introduced in Section 1.  Recall that the defoliation
data contains images of defoliation rates (response) and elevations
(covariate). The images consist of $64\times 64$ evenly spaced pixels
$\textbf{s}_i$s, and therefore, $n=4096$. The response $y(\textbf{s})$
and the covariate $x_1(\textbf{s})$ represent the centered-and-scaled 
defoliation rate and scaled
elevation measured at pixel $\textbf{s}$ respectively 
(as displayed in Figure \ref{fig:1}). We used the
centered-and-scaled $x_1$ as the covariate
surface $x$, i.e.,
$x(\textbf{s})=(x_1(\textbf{s})-\overline{vec(x_1)})/\textrm{std}(vec(x_1))$,
where $vec(x_1)$ denotes the vector of $x_1$ values at the 4096
pixels, and $\overline{vec(x_1)}$ and $\textrm{std}(vec(x_1))$ are the
sample mean and standard deviation of $vec(x_1)$. $J=4$ was used, and
thus, $m=2048$ wavelet coefficients are involved in our model.

We fixed $\mu=\nu=6$ and fit Model I. Prior (1) was placed on $\gamma$
with the Bernoulli probabilities corresponding to resolution levels 0
to 4 being $0.5, 0.5\phi, 0.5\phi^2, 0.5\phi^3$ and $0.5\phi^4$
respectively. We somewhat arbitrarily chose $\phi=0.9$ to
  produce some degree of flatness in the estimates.  A Markov chain
of length 20,000 was simulated from the posterior distribution
$p(\beta,\gamma,\sigma,\tau|\textbf{y},X)$ specified by
(\ref{post:likelihood}) using Algorithm I, and the first half was
treated as burn-ins.  The initial point $\beta^{(0)}$ for the $\beta$
chain was generated from $ N(\hat{\beta},10^{-4}I_m)$, where
$\hat{\beta}$ was chosen as the least squares estimate of $\beta$.  It
took about 2.25 hours to draw 10,000 posterior samples.  Convergence
was assessed by applying Gelman-Rubin factors to 5 parallel Markov
chains.  We also applied the method introduced in Section 3.2 to
classify the pixels.

\begin{figure}[htb]
\begin{center}
\includegraphics[angle=0,scale=1]{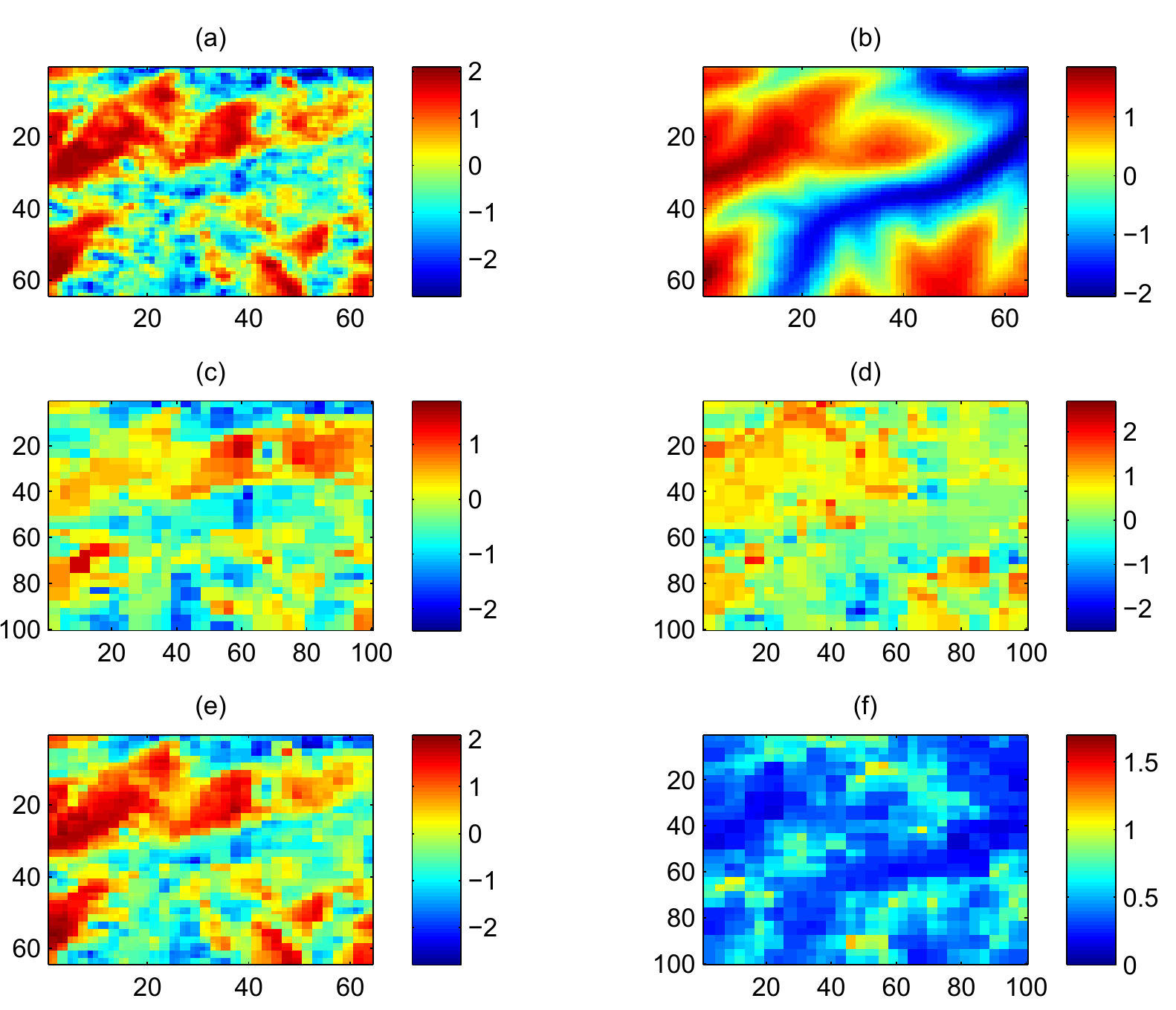}
\caption{\textit{\footnotesize (a): Centered-and-scaled defoliation rate $y$; (b): Centered-and-scaled elevation $x$;
(c): Estimated intercept $\hat{A}$; (d): Estimated slope $\hat{B}$;
(e): Fitted value $\hat{y}$; (f): PSD of the slope $B$. Prior (1) with
$\phi=0.9$ was used.}}
\label{fig:ABSigmaReal}
\end{center}
\end{figure}

Figure \ref{fig:ABSigmaReal} displays the estimated intercept
$\hat{A}$, the estimated slope $\hat{B}$, the fitted defoliation rate
$\hat{y}$ and the PSD of the slope $B$. In particular, the
  images of $\hat{A}$, $\hat{B}$ and the PSD were constructed over a
  $100\times 100$ lattice set of locations in $[0,1)\times [0,1)$ to
  display the posterior samples at new locations; while the $\hat{y}$
  image was constructed over the $64\times 64$ lattice set of
  locations in $[0,1)\times [0,1)$ where the data were drawn allowing
  us to compare $\hat{y}$ with $y$ at the observed locations.  We
observe that $\hat{B}$ is positive at most of the pixels, which shows
an overall positive relationship between the defoliation rate and
elevation.  Furthermore, $\hat{B}$ is slightly smaller at the
locations where the elevation is small.  We also observe that in the
regions where the elevation changes quickly, the PSD of the slope
deviates considerably from zero. Finally, the image $\hat{y}$ appears
to resemble the observed defoliation rate image $y$. 
Our findings on $\hat{B}$ and $\hat{y}$ are similar to those
made by Zhang \textit{et al.} (2011) who used LASSO algorithm to perform
the computations, but again, we are also able to characterize the
uncertainty in the relationships. 

\begin{figure}[htb]
\begin{center}
\includegraphics[height=40cm, width=30cm, angle=0,scale=0.35]{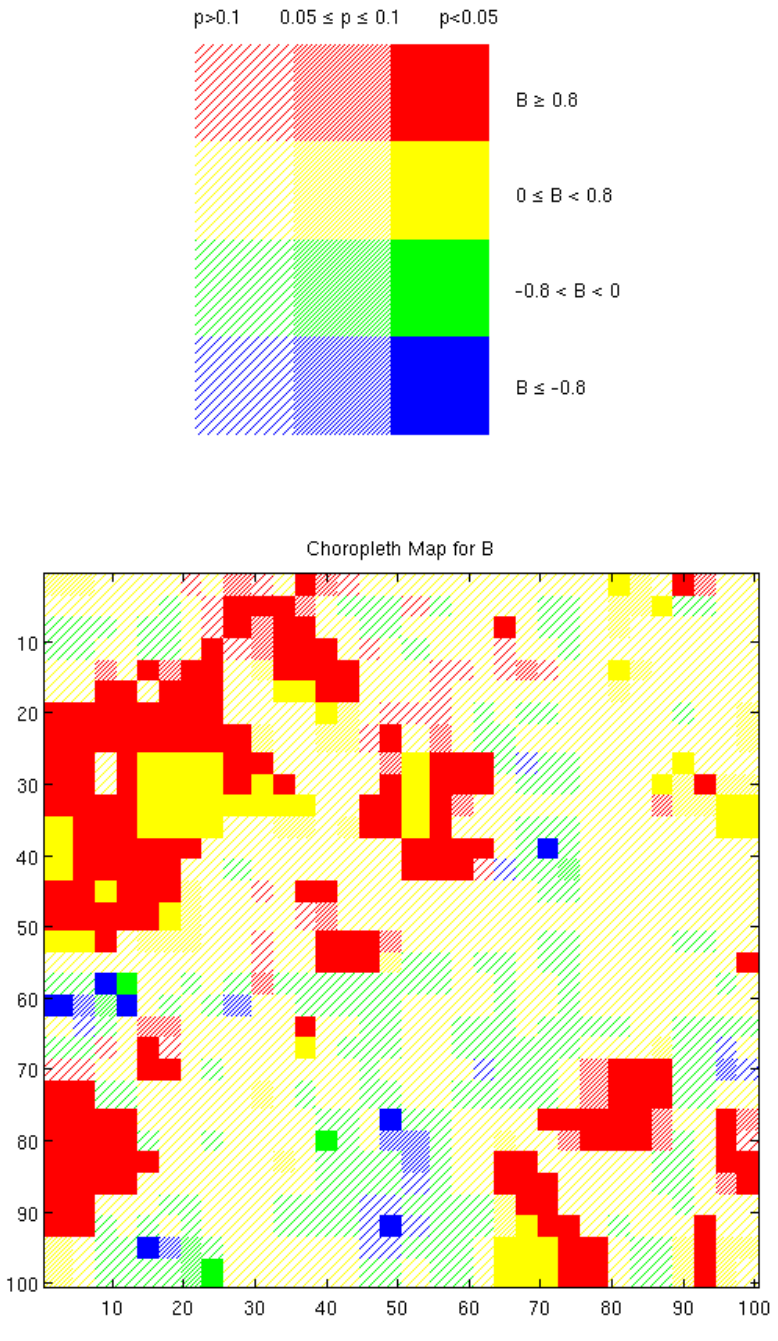}
\caption{\textit{\footnotesize Choropleth map for the slope surface $B$. Prior (1) with
$\phi=0.9$ was used. 
Red indicates $\hat{B}_i\ge0.8$; Yellow: $0\le\hat{B}_i<0.8$; 
Green: $-0.8<\hat{B}_i<0$; Blue: $\hat{B}_i\le-0.8$. 
Filled boxes: $p<0.05$; 
boxes with dense lines: $0.05\le p\le 0.1$; 
boxes with thin lines: $p>0.1$.}}
\label{fig:classification-realdata}
\end{center}
\end{figure}

Figure \ref{fig:classification-realdata} displays the choropleth map
of the slope in which we chose $\Delta=0.8$ (about $1/3$ the maximum
of $|\hat{B}|$).  We observe that in the upper-left region, the
relationship between defoliation rate and elevation is strong and
positive, while in the nearly central region, the relationship between
defoliation rate and elevation is not strong.  We also observe that,
at a small number of locations, $p<0.05$ and $B\le -0.8$ which shows
that the relationship there is strong and negative.

\section{Discussion}

Zhang \textit{et al.} (2011) applied a wavelet approach to transform the spatial
concurrent linear model into a linear model with design matrix induced
by a wavelet structure, and they implemented LASSO to handle the
estimation problem. With their approach, however, it is difficult to conduct inferences
using their method. To address this, we have developed a Bayesian
variable selection approach based on the model proposed by Zhang
\textit{et al.} (2011).  Specifically, we applied a Bayesian model
similar to one proposed by George and McCulloch (1993), in which we
introduced a vector $\gamma$ of Bernoulli variables for the model
coefficients so that the selection and estimation of the nonzero
coefficients can be simultaneously achieved. The proposed approach is
highly flexible and computationally efficient, and should be useful in
many practical situations where the data display complex nonstationary
patterns.  In addition, we developed a Gibbs sampler for posterior sampling
that involves no complicated matrix computation. Hence, this is
efficient for handling relatively large datasets.  Furthermore, as
demonstrated in simulated and real data analysis, our approach is
effective in detecting the spatial locations where the response has a
relationship with a covariate, and provides statistical evidence for
such detections.

We have placed Bernoulli priors on $\gamma$.
Other priors such as Markov chain priors can also be applied by
invoking a tree structure (see Romberg \textit{et al.}, 2001).
The support of any Haar wavelet basis function, which we call a parent, 
is divided into four equal adjacent pieces at the same level, which we call children,
with each piece being the support of a Haar wavelet basis function.
Since any basis support corresponds to a 0-1 variable $\gamma_j$,
we also call $\gamma_{j'}$ the parent of $\gamma_j$ if their
corresponding basis supports have such parent-children relationship.
Following Romberg \textit{et al.} (2001), a Markov chain prior is defined to be
\begin{equation}\label{MCprior}
p(\gamma_j|\gamma_{-j})=p(\gamma_j|\gamma_{j'}),
\end{equation}
where $\gamma_{-j}=\{\gamma_i| i\neq j\}$, and $\gamma_{j'}$ is the parent of $\gamma_j$.
The equation (\ref{MCprior}) means that the distributional properties of a child only depends on its parent.
Let the transition probability be $p(\gamma_j|\gamma_{j'})=p_{\gamma_{j'},\gamma_j}$.
We have numerically examined Markov chain priors with $p_{0,0}=0.9$, $p_{0,1}=0.1$, 
$p_{1,0}=0.1$, $p_{1,1}=0.9$, and found that they did not perform
as well as Bernoulli priors and LASSO when estimating a piecewise constant surface. 
The reason might be that a piecewise constant surface has too much local flatness, and hence, even if
a parent corresponds to a nonzero wavelet coefficient,
its four children may still correspond to zero wavelet coefficients,
which makes the connection between the parent and children weak.
Under such circumstances, Bernoulli priors which assume independence
among the basis functions may be better choices.

Two future extensions of the current work might be also worth
mentioning.  First, Dunson (2009) proposed a nonparametric Bayesian
approach to model the basis coefficients in a longitudinal model. In
his method, the prior distribution of the basis coefficients is
nonparametric; in particular, they used a Dirichlet process prior,
which provides a great deal of flexibility.  Dunson (2009) found that
the nonparametric prior works well for modeling the model
coefficients, and it seems reasonable to extend that work to our
model.

Second, in our model, the coefficients are sparse, and so even if the
the number of parameters is large,  
the estimation results are still satisfactory. Although a sparse 
coefficient vector is common in the regression models associated with wavelets, 
it is still interesting to fit a model with non-sparse coefficients and 
examine the results. One article about the identification of the sparseness 
pattern of the model coefficients is given by Meinshausen and Yu (2009) who examined 
the impact of sparseness on LASSO estimates. There seems to be little literature 
handling this problem under a Bayesian framework, and so we intend to
explore this further in the future.

{\bf Supplement Material:}\,\, Sampler derivations for algorithm I can be found in the first author's website
http://www.stat.wisc.edu/$\sim$shang/

\textbf{Acknowledgments}\,\, The authors would like to thank the Editor, an associate editor and the anonymous 
referees for valuable and constructive suggestions leading
to substantial improvements in the article. 
The authors also thank Bret Larget, Kam Tsui, Jun Zhu, Rick Nordheim and Jun Shao for many useful suggestions
on this paper. 

\noindent{\large\bf References}
\begin{description}
\small
\bibitem{BFV} Brown, P., Fearn, T. and Vannucci, M. (2001).  Bayesian Wavelet Regression on Curves With Application to a Spectroscopic
Calibration Problem. \textit{Journal of the American Statistical Association,} \textbf{96}, 398--408.

\bibitem{BVF} Brown, P., Vannucci, M. and Fearn, T. (2002). Bayes Model Averaging with Selection of Regressors. \textit{Journal of the Royal Statistical Society, Series B,} \textbf{64}, 519--536.

\bibitem{CGMM} Casella, G., Gir\'{o}n, F. J., Mart\'{i}nez, M. L. and Moreno, E. (2009).  
\newblock Consistency of Bayesian Procedures for Variable Selection. 
\newblock{\em The Annals of Statistics,} {\bf 37}, 1207--1228.

\bibitem{CKM} Chipman, H., Kolaczyk, E. and McCulloch, R. (1997). Adaptive Bayesian Wavelet Shrinkage. 
\textit{Journal of the American Statistical Association,} \textbf{92}, 1413--1421.

\bibitem{CG} Clyde, M. and George, E. (2000). Flexible Empirical Bayes Estimation for Wavelets. \textit{Journal of the Royal Statistical Society, Series B,} \textbf{62}, 681--698.

\bibitem{CPV} Clyde, M., Parmigiani, G. and Vidakovic, B. (1998). Multiple Shrinkage and Subset Selection in Wavelets. \textit{Biometrika,} \textbf{85}, 391--401.

\bibitem{CNB} Crouse, M. S., Nowak, R. D. and Baraniuk, R. G. (1998). Wavelet-Based Statistical Signal Processing Using Hidden
Markov Models. \textit{IEEE Transactions on Signal Processing,} \textbf{46}, 886--902.

\bibitem{DA} Daubechies, I. (1992). \textit{Ten Lectures on Wavelets.} CBMS-NSF Regional Conference Series in Applied Mathematics 61.

\bibitem{D} Dunson, D. (2009). Nonparametric Bayes Local Partition Models for Random Effects. 
\textit{Biometrika,} \textbf{96} 249--262.

\bibitem{FWFA} Fan J., Wu, Y. and Feng, Y. (2010). Local Quasi-likelihood With a Parametric Guide. \textit{The Annals of Statistics,} \textbf{37}, 4153--4183.

\bibitem{FLS} Fern\'{a}ndez, C., Ley, E., and Steel, M. F. (2001). Benchmark Priors for Bayesian Model Averaging. \textit{Journal of Econometrics,} \textbf{100}, 381-427.
 
\bibitem{GCSR} Gelman, A., Carlin, J. B., Stern, H. S. and Rubin, D. B. (2003). \textit{Bayesian Data Analysis} (2nd ed). Chapman $\&$ Hall/CRC.
 
\bibitem{GM} George, E. and McCulloch, R. (1993). Variable Selection via Gibbs Sampling. \textit{Journal of the American Statistical Association,} \textbf{88}, 881--889.

\bibitem{GR} Godsill, J. S. and Rayner, P. J. W. (1998). Robust Reconstruction and Analysis of Autoregressive Signals in Impulsive
Noise Using the Gibbs Sampler. \textit{IEEE Transactions on Speech and Audio Processing,} \textbf{6}, 352--372.

\bibitem{J} Jiang, W. (2007). Bayesian Variable Selection for High Dimensional Generalized Linear Models: Convergence Rates of the
Fitted Densities. \textit{The Annals of Statistics,} \textbf{35}, 1487--1511.

\bibitem{JT} Jiang, W. and Tanner, M. (2008). Gibbs Posterior for Variable Selection in High-Dimensional Classification and Data Mining. 
\textit{The Annals of Statistics,} \textbf{36}, 2207--2231.

\bibitem{KM} Kleiner, K. and Montgomery, M. (1994). Forest Stand Susceptibility to the Gypsy--Moth (lepidoptera, lymantriidae)--Species and Site Effects on Foliage Quality to Larvae. \textit{Environmental Entomology,} \textbf{23}, 699--711.

\bibitem{LPMCB} Liang, F., Paulo, R., Molina, G., Clyde, M. and Berger, J. (2008). Mixtures of $g$-Priors for Bayesian Variable
Selection. \textit{Journal of the American Statistical Association,} \textbf{103}, 410--423.

\bibitem{LZ} Li, F. and Zhang, N. R. (2010). Bayesian Variable Selection in Structured High-Dimensional Covariate Spaces with Applications in Genomics. \textit{Journal of the American Statistical Association,} \textbf{105}, 1202--1214.

\bibitem{MY} Meinshausen, N. and Yu, B. (2009). LASSO-type Recovery of Sparse Representations for High-Dimensional Data. 
\textit{The Annals of Statistics,} \textbf{37}, 246-270.

\bibitem{MGC} Moreno, E., Gir\'{o}n, F. J. and Casella, G. (2010). 
\newblock Consistency of Objective Bayes Factors as the Model Dimension Grows. 
\newblock{\em The Annals of Statistics,} \textbf{38}, 1937--1952.

\bibitem{NG} Nott, D. and Green, P. (2004). Bayesian Variable Selection and Swendsen-Wang Algorithm. 
\textit{Journal of Computational and Graphical Statistics,} \textbf{13}, 141--157.

\bibitem{RCB} Romberg, J., Choi, H. and Baraniul, R. (2001). Bayesian Tree-Structured Image Modeling Using Wavelet-Domain Hidden
Markov Models. \textit{IEEE Transactions on Image Processing,} \textbf{10}, 1056--1068.

\bibitem{SC} Shang, Z. and Clayton, M. K. (2010). Consistency of Bayesian Model Selection for Linear Models
With A Growing Number of Parameters. {\it Journal of Statistical Planning and Inference,} in press.

\bibitem{SF} Smith, M. and Fahrmeir, L. (2007). Spatial Bayesian Variable Selection With Application to Functional Magnetic Resonance
Imaging. \textit{Journal of the American Statistical Association,} \textbf{102}, 417--431.

\bibitem{TEW} Townsend, P. A., Eshleman, K. N. and Welcker, C. (2004). Remote Sensing of Gypsy Moth Defoliation to Assess Variations in Stream Nitrogen Concentrations. \textit{Ecological Applications,} \textbf{14}, 504--516.

\bibitem{W} Wheeler, D. C. (2009). Simultaneous coefficient penalization and model selection in
geographically weighted regression: the geographically weighted lasso. \textit{Environment and
Planning A} \textbf{41}, 722--742.

\bibitem{WW} Wheeler D. C., and Waller L. A. (2009). Comparing spatially varying coefficient models: a case
study examining violent crime rates and their relationships to alcohol outlets and illegal
drug arrests. \textit{Journal of Geographical Systems} \textbf{11},  1--22.

\bibitem{WN} Wolfe, P., Godsill, S. and Ng, W. (2004). Bayesian Variable Selection and Regularization for Time-Frequency Surface
Estimation. \textit{Journal of the Royal Statistical Society, Series B,} \textbf{66}, 575--589.

\bibitem{YL1} Yuan, M. and Lin, Y. (2005). Efficient Empirical Bayes Variable Selection and Estimation in Linear Models.
\textit{Journal of the American Statistical Association,} \textbf{100}, 1215--1225.

\bibitem{ZCT} Zhang, J., Clayton, M. K. and Townsend, P. A. (2011). Functional Concurrent Linear Regression Model for Spatial Images. 
\textit{Journal of Agricultural, Biological and Environmental Statistics,} {\bf 16}, 105--130.

\end{description}

\newpage

\newpage

\newpage

\newpage

\newpage
\end{document}